\documentclass[twocolumn,aps,pre,showpacs,floatfix]{revtex4}
\usepackage{epsfig}
\usepackage{amsmath}
\usepackage{psfig}
\usepackage{amssymb}
\newcommand{\dd}{\text{d}}
\newcommand{\sR}{\text{\tiny R}}
\newcommand{\ee}{\text{e}}

\newcommand{\p}{\partial}
\newcommand{\bx}{\text{\bf x}}

\newcommand{\eps}{\varepsilon}

\newcommand{\bq}{\text{\bf q}}

\begin{document}
\noindent pr\'epublication LPT Orsay 03-30; preprint ITP-UU-03/19 

\title{INFINITELY-MANY ABSORBING-STATE NONEQUILIBRIUM PHASE TRANSITIONS}
\author{Fr\'ed\'eric van Wijland$^{1,2}$}

\affiliation{${}^1$Institute for Theoretical Physics, Postbus 80006, 3508 TA
Utrecht, The Netherlands}

\affiliation{${}^2$Laboratoire de physique th\'eorique, b\^atiment 210,
Universit\'e de Paris-Sud, 91405 Orsay cedex, France.}

\date{\today}
\begin{abstract}
We present a general field-theoretic strategy to analyze three connected
families of
continuous phase transitions which occur in nonequilibrium steady-states. We
focus on transitions taking place between an active state and one
aborbing state, when there exist an infinite number of such absorbing
states. In such transitions the order parameter is coupled to an auxiliary
field. Three situations arise according to whether the auxiliary field is
diffusive and conserved, static and conserved, or finally static and not
conserved.
\end{abstract}

\maketitle

\section{The ubiquity of absorbing-state transitions}
This overview is devoted to a study of
nonequilibrium phase transitions taking place between the active and the
absorbing state of a
system, as some control parameter is varied across a
threshold value. Such transitions are
encountered in a variety of fields ranging from chemical kinetics to the
spreading of computer viruses~\cite{reviews}. From a theoretical standpoint
absorbing state transitions form natural
counterparts to equilibrium phase transitions. The transition rates used in the
stochastic dynamics employed to model the
physical phenomenon under consideration do not satisfy detailed balance (with
respect to an {\it a priori} defined  energy function). In spite of
this apparent freedom, the number of universality classes that the transition
can fall into is incredibly small. Among known universality classes, that of
directed percolation (DP) is by far the broadest. And indeed, in the absence of
additional symmetries or conservation laws, as was conjectured twenty years ago
by Grassberger~\cite{grassberger79}, an absorbing state transition will invariably fall into the DP
universality class. The interest in absorbing state transitions was further
enhanced as Dickman and coworkers~\cite{fes} established a
one-to-one correspondence with self-organized critical systems (see
\cite{jensen98} for a review on self-organized criticality). They were able to
show that
the scaling behavior observed there was entirely governed by an underlying
nonequilibrium phase transition (which, as a side effect, somewhat tempers the mystics of SOC).\\

The study of exactly which microscopic ingredients make an absorbing state transition
{\it not} belong to the DP class has almost grown into a field of its own. It
was early realized that if the microscopic dynamics possesses additional
conservation laws the universality class of the transition could be different.
Discrete conservation laws, such as the conservation of the parity of the number
of particles~\cite{taubercardy}, are known to be driving
the transition to an independent universality class (the Parity Conserving or
Voter class~\cite{dornicchatechavehinrichsen}). A recent study attempts to
provide a comprehensive table of all possible transitions involving the dynamics
of a single field~\cite{chatekockelkoren}.
Besides, the effect of a continuous symmetry was
shown either to change the universality class of the transition~\cite{kree,
wij98} or to simply destroy its continuous nature~\cite{wij2000}. The continuous
symmetry present in the systems studied there arose from a local conservation
law.\\

An independent direction of research has focused on absorbing-state transitions
in which the order parameter freezes into one among an infinity of absorbing
states, but without any additional conservation law. The paradigmatic example of
a system showing such a behavior is the pair contact process, initially
introduced by Jensen and Dickman~\cite{jensen93b}, for which Mu\~noz and
coworkers~\cite{munozetal} devised a convincing phenomenological picture that we
shall later rely on for our analytic manipulations. 
The sensitivity of the DP class with respect to the coupling to an auxiliary
field might actually provide a way out for explaining the difficulty of
experimentally
observing of the DP class~\cite{hinrichsen-exp}, in spite of
recent efforts (see \cite{rupprichterrehberg} and references therein). 
Among the few effects that may lead a transition to depart from the directed
percolation class we list quenched
disorder and the coupling to an auxiliary field. The present works focuses on the
latter (which was first formalized in those terms by Grassberger, Chat\'e and Rousseau~\cite{chate}). We should also emphasize that the following exposition is a one-sided
approach to those phenomena, relying solely on the field-theoretic approach, thus
completely omitting the huge numerical efforts invested in the field.\\

The existence of an infinite number of absorbing states (in the large-system
limit) and the coupling to an auxiliary field are
the common characteristics to the microscopic models that we wish
to investigate here. We will
provide a full renormalization group picture of the phase transition at work in
systems possessing an infinite number of absorbing states, with or without an
additional conservation law. We shall
rely on a combination of exact mappings and phenomenological Langevin equations
approach as a starting point for the calculations. We begin by introducing the
three families of models the critical behavior of which we wish to understand.
Then we sketch the field-theoretic stragegy that we will follow on the simplest
example of an epidemic. As we turn our attention to more complex models we shall
identify which are the new key ingredients that pose technical difficulties. By
doing so we come across an unexpected connection between the last two families
of models we want to consider. The conclusion section lists some open problems
with the present approach and gives directions for possible future works.\\

\section{Three families of processes}
\subsection{Spreading of an epidemic (SIS)}
A model for the spreading of an epidemic which is well-knwon to epidemiolgists
is the so-called Susceptible-Infected-Susceptible model (SIS). The population is
divided into two groups, the susceptible $A$'s and the infected $B$'s, whose
interactions are contamination of an $A$ by a $B$ upon encounter,
\begin{equation}\label{contamination}
A+B\to B+B
\end{equation}
and spontaneous recovery of an infected individual,
\begin{equation}\label{desintegration}
B\to A
\end{equation}
The motion of the individuals will be assumed to be diffusive, with diffusion
constants $D_A$ and $D_B$ for the susceptible and the infected individuals,
respectively. While the assumption of diffusive motion could itself be discussed
(the very same reaction processes with underlying motion provided by a chaotic
flow are used to model plancton population dynamics in the ocean~\cite{tel}) we
will take it for granted. The hope is that this simple assumption will apply to the
majority of systems with such competitive rules.\\

When the average density of individuals $\rho$, which is obviously conserved by the above
rules, is high enough, the infection survives indefinitely. At low density, on
the contrary, the epidemic becomes extinct exponentially fast. And there exists
a critical density threshold $\rho_c$ separating the two steady-states, {\it i.e.} the
active one, with ongoing spreading, from the absorbing one in which the extinct
epidemic cannot be revived. In the absorbing states, $A$ particles are freely
diffusing.\\

The critical properties of, say the density of infected individuals, in the
vicinity or at the density threshold were studied both analytically~\cite{kree,
wij98, wij2000} and
numerically~\cite{freitaslucenasilvahilhorst1,fulcomessiaslyra,perlsmanhavlin},
with a rich variety of behaviors. There is even still some controversy over the order of
the transition in low space dimensions (mean-field predicts a continuous
transition).

It is not hard to coin a mean-field phenomenological evolution equation for the order
parameter field $\psi$,
\begin{equation}\label{mean-field-SIS}
\p_t\psi=D_B\Delta \psi+(\rho/\rho_c-1)\psi-g_1\psi^2
\end{equation}
where $g_1$ is a coarse-grained contamination rate. From this equation we deduce
that the
density of infected individuals $\psi$ in the steady-state, undergoes a continuous transition from an
active state in which, as $\rho\to\rho_c^+$, $\psi\sim(\rho-\rho_c)^\beta$, to an
absorbing state as $\rho\leq \rho_c$, with $\psi=0$. At the mean-field level
$\beta=1$. Similarly the relaxation rate to the steady-state occurs over a
typical time scale $\tau\sim \xi^z$ with $z=2$ and there is typical correlation
length $\xi\sim
|\rho-\rho_c|^{-\nu}$, with $nu=1/2$. At the critical point the density decays
as $\psi(t)\sim t^{-\delta}$ with $\delta =1$. 
This set of mean-field exponents will accurately describe the scaling properties
of the transition whenever the space dimension is high enough to allow diffusion
to quickly homogeneize space fluctuations. However, in low space dimensions, where
random walks are recurrent or weakly transient, correlations between long-lasting fluctuations will play a role and
the scaling behavior will be modified. In a region of linear size $\ell$ density
fluctuations are of order $\ell^{d/2}$, and thus relative density fluctuations
are of the order $\ell^{-d/2}$. From (\ref{mean-field-SIS}) we see that the
reaction relaxation time goes
roughly inversely proportional to the density fluctuations, hence the time it
takes for the reaction to equilibrate over the region of size $\ell$ is of the
order $\ell^{d/2}$. But this mean-field reasoning holds only if diffusion has
acted fast enough so as to erase space fluctuations over the domain of size
$\ell$, which takes a time of the order $\ell^2$, and for mean-field to hold we
must have $\ell^2\ll \ell^{d/2}$, that is $d>4$.\\

The features which make of the SIS a system whose behavior is not of DP type are
the following: the order parameter field is coupled to an auxiliary conserved
field $\phi$ standing for the local density fluctuation of particles (independently of their
$A$ or $B$ type). There exist an infinite number of absorbing states, but
admittedly all of them are equivalent. In \cite{grassberger-discussion}
Grassberger makes a distinction between fully ergodic absorbing states (such as
the ones resulting from the epidemic becoming extinct) and those
which are frozen in. We now turn to an example of the latter.
\subsection{Fixed Energy Sandpiles (FES)}
The second family of systems that we would like our analysis to encompass has,
at first sight, no relationship with the above. Stochastic fixed energy
sandpiles (FES), known as the Manna model~\cite{manna,fes,bigfes}, are defined as follows. Grains of
sand are distributed on the sites of a lattice. Whenever a site is occupied by
more than two grains, these are randomly distributed to the nearest neighbors.
After a while, at sufficiently low grain density, all the sites will have less
than two grains and the activity remains frozen in forever. At high grain
density the toppling processes will keep occurring throughout the system, thus
leaving a finite density of active --or toppling-- sites in the infinite time
limit. Of course, the system is supposed to be closed (the grains cannot escape
though the boundaries). Identify now an active site with an infected individual
$B$ and a still site with a susceptible $A$. Only the $B$'s are now diffusing,
but otherwise the reaction rules (\ref{contamination},\ref{desintegration}) continue
to roughly describe what is taking place in the system. Interestingly, and this
is the important contribution of Vespignani and coworkers, a very large number
of cellular automata can effectively be described by the SIS model in which the
susceptibles are static. While this slight variation on the model present in the
previous paragraph sounds innocuous, it will prove a dramatic change for te
scaling properties. To get a
first feeling of why it is so, we note that, as previously, the order parameter
is coupled to a conserved field. But now this field is static, or rather its
dynamics is slaved to that of the order parameter. We may expect, thus, that the
fluctuations of this field will introduce memory effects, and lead to increased
deviations to mean-field behavior. Again there exist an infinite number of
absorbing states, but we do not know a priori what their distribution is.
Pastor-Satorras and Vespignani~\cite{pastor} performed the first study of the
epidemic with static healthy individuals, after the general characteristics of
systems described by those schematic reaction rules had been
identified by Rossi {\it et al.}~\cite{rossi} by studying two distinct automata,
the Activated Random Walkers and the Conserved Transfer Threshold Process.

\subsection{Pair Contact Process (PCP)}
The third and final family of models we will bend over is embodied by the Pair
Contact Process (PCP), a reaction involving a single species of static
particles denoted by $C$. These excluding particles may either annihilate when on neighboring sites according
to $C+C\to\emptyset$ or produce an offspring $C+C\to C+C+C$. But they do not
diffuse. The order parameter
of the transition is the density of pairs of nearest neighbor particles. At high
branching rate the steady-state  exhibits a finite density of particles, while
at low branching the system eventually settles in a frozen state in which all
particles are isolated. The order parameter of the transition is the local pair
density. The connection between the PCP and the above two
processes lies in the identification of a pair with a $B$ particle, and of an
isolated particle with an $A$. Unfortunately there is no ready-to-use one-to-one mapping
between a configuration of $A$'s and $B$'s and a configuration of the $C$'s.
One can draw a heuristic correspondence based on physical intuition, which can be written in a rather loose notation in a form
reminiscent of the previous paragraphs:
\begin{equation}
A+B\to B+B,\;\;B\to \emptyset,\;\;B\to A
\end{equation}
While there is obviously no particle conservation one may easily identify a
branching process similar to the contamination one and two growth-limiting
processes. Note also that the branching of a pair induces effective diffusion
for the $B$'s.  A large number of systems, such as the dimer
reaction~\cite{jensen93b},
the dimer-dimer~\cite{many} and dimer-trimer~\cite{kohlerbenavraham} reactions, or the threshold
transfer process~\cite{mendes94} are modeled by the above
mechanism.
\section{Field-theoretic techniques: a unified treatment}
\subsection{Action for the epidemics}
There are well-established techniques~\cite{ft} for mapping a reaction-diffusion process
in which particles diffuse, branch, annihilate, and possibly exclude each other,
onto a field-theory that {\it exactly} encodes the microscopic dynamics. We will
not review how such a mapping is achieved. It suffices to know that the resulting
field theory features, for each species of particles, a density field, the first
moment of which gives the local density of particles, and  a response field
(usually bearing an overbar) with no straightforward physical interpretation.
Denoting by $\psi$ the order parameter field and by $\phi$ the local density
fluctuation we have the following action
\begin{equation}\label{actionSIS}\begin{split}
S[\bar{\psi},\psi,\bar{\phi},\phi]=\int\dd^d x\dd
t\Big[&\bar{\psi}(\p_t+\sigma-D_B\Delta)\psi\\
&+\bar{\phi}(\p_t-D_A\Delta)\phi\\&
+\bar{\phi}(D_A-D_B)\Delta\psi\\
&+\bar{\psi}\psi^2-\bar{\psi}^2\psi+\bar{\psi}\psi(\bar{\phi}+\phi)\Big]
\end{split}\end{equation}
where we have omitted higher order terms in the fields and have not specified
any names for the interaction vertices. Note that the free field
$\bar{\phi},\phi$ could be integrated out to yield an effective action for
the order parameter $\psi$ and its response field $\bar{\psi}$ alone. The
renormalization group analysis of (\ref{actionSIS}) was
performed~\cite{kree,wij98,wij2000} with the following results~: the coupling of
the order parameter to an auxiliary diffusive and conserved field drives the
transition to two universality classes different from the DP one according to
whether $0<D_A<D_B$ or $D_A=D_B$. Even more surprisingly, the continuous phase
transition is outweighed by a first order one below the upper critical dimension
when $0<D_B<D_A$. While a first-order transition was indeed
observed in
a two-dimensional system, it has been both ruled
out~\cite{fulcomessiaslyra,hinrichsen} and confirmed~\cite{barratvespignani} in
one dimension. In the following table we recall the results for the critical
exponents given to first order in $\eps=4-d$ (a star superscript indicates a
result holding to all orders in $\eps$) and provide for comparison the
directed percolation expression~\cite{janssenDP}. The exponents $\eta$ and
$\bar{\eta}$ are the anomalous dimensions of the $\psi$ and $\bar{\psi}$ fields,
respectively, defined as $\psi_{\sR}\sim \ell^{-\frac{d+\eta}{2}}$ and
$\bar{\psi}_{\sR}\sim \ell^{-\frac{d+\bar{\eta}}{2}}$, where $\ell$
is a length scale.
\begin{equation}\label{table}
\begin{array}{|c|c|c|c|}
\hline
\text{exponent}&D_A=D_B&D_A<D_B&\text{DP}\\
\hline
z&2^*&2^*&2-\frac{\eps}{12}\\
\hline
\nu^{-1}&2-\frac{\eps}{2}{}^*&2-\frac{\eps}{2}{}^*&2-\frac{\eps}{4}\\
\hline
\eta&-\frac{\eps}{8}&0^*&-\frac{\eps}{6}\\
\hline
\bar{\eta}&-\frac{\eps}{8}&-0.313\eps&-\frac{\eps}{6}\\
\hline
\beta=\nu\frac{d+\eta}{2}&1-\frac{\eps}{8}&1^*&1-\frac{\eps}{6}\\
\hline
\end{array}
\end{equation}
It is worth commenting on Table~(\ref{table}): several exponents appear to be
given to all orders in $\eps$. That the $z$ exponent takes the value 2 is
understable since the conserved and freely diffusive field $\phi$ imposes its
(slow) relaxation scale. Hence the superdiffusive behavior observed in DP ($z<2$
) when it is not coupled to any auxiliary field is overcome by the slowest
relaxing modes, which happen to be freely diffusive modes (with diffusion
constant $D_A$). This is a very robust property that can be seen to hold even
if we took the omitted quartic terms in
(\ref{actionSIS}) into account. It is rooted in the absence of renormalization of
the $\phi$ propagator. The exactness of the correlation length exponent
$\nu=2/d$ has a very different origin. A shift of the density field $\phi$ by a
constant has the same effect as shifting the $\psi$ field mass term. This
implies a exact identity between vertex function, which carries over to the
renormalized quantities, thus yielding an additional relationship between
exponents, namely that $\nu^{-1}$ equals the scaling dimension of $\phi$, which is
$d/2$. Unfortunately this continuous symmetry is broken by quartic terms whose
effect in low-dimensions is ill-controled. This has led to some debate in the
literature~\cite{janssencomment,freitaslucenasilvahilhorst2,fulcomessiaslyra}.
Quite understanbly, the exact knowledge of some exponents (if blindly taken for
granted) eases numerical
analysis and yields higher precision results on the remaining exponents.

\subsection{One-loop-expansion for FES}
\begin{figure}[htb]
  \centerline{ \epsfxsize=5.2cm \epsfbox{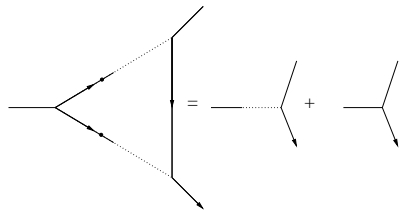} }
   \vspace{0.cm}
\caption{The one-loop diagram combines the $g_4$ and $g_2$ vertices to yield
effective $g_3$ and $g_1$ couplings. A black dot on a leg means that the vertex
is proportional to the square of the momentum flowing through the leg. The
dotted line means that the dangling leg is at time earlier than the other two
making up the vertex.}
\label{fig1}
\end{figure}

\begin{figure}[htb]
  \centerline{ \epsfxsize=5.2cm \epsfbox{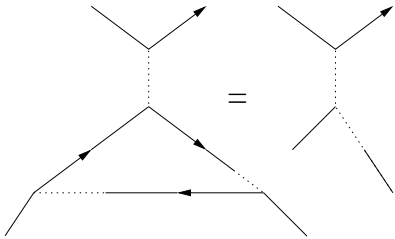} }
   \vspace{0.cm}
\caption{Two $g_3$ vertices are combined with a $g_5$ vertex to yield an
effective $\int\dd^dx \dd t\;\bar{\psi}\psi\left[\int^t\dd t'\psi(t')\right]^2$
interaction.}
\label{fig2}
\end{figure}

One crucial difference between the SIS and the FES is that the $A$ particles
become static, which can be achieved simply by setting $D_A=0$ in
(\ref{actionSIS}). Note that an immediate consequence of setting the auxiliary
field diffusion constant to zero is that it ceases to impose its own dynamic
scale (in particular, we should now expect that the order parameter field $\psi$ will
indeed exhibit superdiffusive behavior, and that $\phi$ will follow).  The total density field $\phi$ has its dynamics slaved to the
local fluctuations in the order parameter $\psi$, as can be see by writing the
equation of state for $\phi$:
\begin{equation}\p_t\langle\phi\rangle=D_B\Delta\langle\psi\rangle\end{equation}
This means that the modes of $\phi$ describing the short-scale fluctuations
verify $\phi(\bq,t)\sim-\bq^2\int^t\dd \tau\psi(\bq,\tau)$. When integrating
out the $\bar{\phi},\phi$ fields from the action (\ref{actionSIS}) one is left
with effective interaction of the form
\begin{equation}
g_4\int\dd^d x\dd t\bar{\psi}\psi(\bx,t)\int^t\dd t'\Delta_{\bx}\psi(\bx,t')
\end{equation}
But at small distances (large $\bq$'s) there is no difference between the $g_4$ vertex and
\begin{equation}
g_3\int\dd^d x\dd t\bar{\psi}\psi(\bx,t)\int^t\dd t'\psi(\bx,t')
\end{equation}
Thus it is not surprising that a $g_3$-like vertex is generated already at one
loop, as
can be seen on the Feynman diagram depicted in Fig.~\ref{fig1}.
The fact that such terms are generated would seem to be in conflict with the
local
conservation law. But there is no contradiction as other terms are generated, which will turn out to be irrelevant
from the renormalization group point of view, yet they will take care of preserving this
essential property. We now rewrite the effective action for the order parameter field only:
\begin{equation}\label{actionFES}\begin{split}
S[\bar{\psi},\psi]=\int\dd^d x\dd
t\Big[&\bar{\psi}(\p_t+\lambda\tau-\lambda\Delta)\psi\\
&g_1\bar{\psi}\psi^2-g_2\bar{\psi}^2\psi\\
&+g_3\bar{\psi}\psi\int\dd t'\psi(\bx,t')\\&-g_4\bar{\psi}\psi\int\dd
t'\Delta\psi(\bx,t')\\&-g_5\bar{\psi}\psi\int\dd t'\bar{\psi}\psi(\bx,t')\Big]
\end{split}\end{equation}
The $g_1$ and $g_2$ vertices alone make up the directed percolation action. In fact, in all members of the FES family, vertices of
the form $\int\dd^dx \dd t\;\bar{\psi}\psi\left[\int^t\dd t'\psi(t')\right]^n$ are generated at
one loop. To illustrate this state of affairs, in Fig.~\ref{fig2} we draw the one-loop diagram
generating the $n=2$ term.
In the next section it is explained how to make physical sense of the action
(\ref{actionFES}) and to extract physical quantities from it.
\subsection{The special PCP case}
In the PCP the microscopic dynamics is quite different.
Following the suggestion of Mu\~noz~\cite{munozetal}, it is not hard to see that the directed
percolation action has to be supplemented by an additional
interaction term of the form
\begin{equation}\label{additional}
\delta S_{\text{PCP}}[\bar{\psi},\psi]=-\int\dd^d x\dd
t\bar{\psi}\psi(\bx,t)\left[\ee^{-\lambda\int_0^t\dd t'\;\psi(\bx,t')}-1\right]
\end{equation}
Note that subtracting $-1$ in the bracket in the right-hand-side of
(\ref{additional}) amounts to shifting the mass term of the field theory (which
now vanishes at the mean-field critical point). The mass term being positive, this means
that we are working in the absorbing phase. In other words, the order parameter
relaxes exponentially fast to zero. We choose to perform our analysis with a
positive mass, and we shall carry the necessary renormalizations in the absorbing
phase, as is done in all previous renormalization group studies of absorbing state
transitions. Working in the active phase would lead to the same renormalization
factors, but would imply  following a much more complicated path with exactly the
same outcome.

One could believe that the present exponential
term will not contribute to determining the anomalous scaling properties of the
phase transition at work by arguing as follows. The order parameter will tend to
a constant value and hence this term will be suppressed exponentially fast. But
this reasoning involves fixing the mass and letting time to infinity. We
are interested in the reverse limit in which time is much shorter than any
typical critical time so that the critical fluctuations can develop (but of
course much larger than microscopic time scales). And those limits do not
commute. All the terms making up the
series expansion of $\delta S_{\text{PCP}}$ are relevant at the Gaussian fixed point (the $n$'s power of
$\lambda$ has bare dimension
$2+n\eps$, which indicates that those
terms will constitute relevant perturbation of the directed percolation ${\cal O}(\eps)$ fixed point.

The nature of short-time and short-distance singularities (UV divergencies) dictates the
universality class that a phase transition falls into. This is quite
counter-intuitive since a phase transition is a large-scale collective
phenomenon. And indeed renormalization is not but a refined coarse-graining
procedure which integrates out the
short-time and short-distance degrees of freedom, and which will carry these
pieces of information over to the macroscopic degrees of freedom (the global
order parameter, for instance).
\subsection{Analysis}
In order to study UV divergencies we follow the strategy outlines in \cite{fred}
and we expand $\delta S_{\text{PCP}}$ in powers of
$\lambda$. Then a quick inspection at the one-loop graphs tells us that the
new upper critical dimension is $d_c=6$, below which the theory is
super-renormalizable, and above which mean-field applies. From here on the
notation $\eps$ stands for the deviation of the space dimension with respect to the new upper
critical dimension $d_c=6$: $\eps=6-d$. It is instructive to
write $\delta S_{\text{PCP}}$ in the form
\begin{equation}\begin{split}
\delta S_{\text{PCP}}[\bar{\psi},\psi]=&\lambda\int\dd^d x\dd
t\bar{\psi}\psi(\bx,t)\int^t_0\dd t'\psi(\bx,t')\\&-
\int\dd^d x\dd
t\bar{\psi}\psi(\bx,t)\times\\&
\left[\ee^{-\lambda\int_0^t\dd t'\;\psi(\bx,t')}-1+\lambda\int^t_0\dd t'\psi(\bx,t')\right]
\end{split}\end{equation}
because now the brackets in the right-hand-side contain terms which are
irrelevant in the renormalization group sense. A word of caution is needed here:
the unfortunate wording {\it irrelevant} does not mean those terms are
irrelevant physics-wise, and indeed by throwing them away we would simply lose the
physical mechanism at work in the PCP (there would simply be no more phase
transition). This merely expresses that those additional terms do
not introduce corrections to the large scale effective couplings. There
are many instances of systems described by a field theory in which RG-wise
irrelevant terms govern the phase diagram, but not the anomalous scaling properties, both
in equilibrium (see the review by Amit and Peliti~\cite{AmitPeliti}) and out of
equilibrium (see Janssen and Schmittmann~\cite{JanssenSchmittmann} for the example of a driven
diffusive system or Deloubri\`ere {\it al.}~\cite{olivieruwefred}, more
recently, for providing the analysis of the Pair Contact Process
with Diffusion). All powers of $\lambda$ higher than two therefore constitute
dangerously irrelevant terms.\\

However, having in mind that now the upper critical dimension is shifted up to $d_c=6$, we see
that also $g_1$ is a dangerously irrelevant coupling  to the extent that, as
shown by Mu\~noz and coworkers~\cite{munozetal}, it eventually controls the dynamics of the order parameter by
playing the r\^ole of the leading nonlinear growth-limiting coupling (at late
times).\\

There are now two separate problems. The first one is to renormalize the
field-theory with the $g_5\bar{\psi}{\psi}\int^t\dd t'\psi(t')$ vertex. But this
has already being done almost twenty years ago by Janssen~\cite{JanssenDyP} and
Grassberger and Cardy~\cite{cardygrassberger} (this is the same
field theory that describes dynamical percolation). And then one must
compute the scaling dimension of the dangerously irrelevant operators at the
dynamical percolation fixed point. The second issue is how to extract the
scaling behavior of physical observables knowing that the usual scaling
assumption breaks down. Again it may look surprising that the universality class
is that of dynamical percolation, a process with no absorbing state transition.
This is because in the PCP the phase transition is driven by irrelevant
couplings (which are absent in dynamical percolation).\\

The scaling dimension of $g_1$ at the dynamical percolation fixed point is found
to be $y_{g_1}=-2-\frac{\eps}{7}$. At scale $b\gg 1$, the effective $g_1(b)$ behaves as
$b^{y_{g_1}}$. And then
we use the mean-field scaling function's dependence in $g_1$ (which depends on
$g_1$ as $1/g_1$) which is correct to first order in $\eps=6-d$. The conclusion
is that
\begin{equation}
\langle\psi\rangle\sim b^{-\frac{d+2+\eta}{2}}{\cal
F}(b^{1/\nu}|\sigma|,b^{-z}t,g_1(b))
\end{equation}
with ${\cal F}(x,y,z)\stackrel{z\to 0}{\propto}\frac{F(x,y)}{z}$ to leading
order in $\eps$. We can therefore extract the following critical exponents
\begin{equation}
\beta=1-\frac{3}{14}\eps,\;\;\delta=1-\frac{1}{4}\eps
\end{equation}
which are valid to first order in $\eps$. We have used the dynamical percolation
expressions for the correlation time and correlation length exponents,
$z=2-\frac{\eps}{6}$
and $\nu^{-1}=2-\frac{5}{21}\eps$.

\section{Conclusions}
One of the major drawbacks of the analysis presented here, as far as the FES and
PCP-like systems are concerned, is the absence of
agreement with numerical simulations. That there is disagreement with the
dynamical percolation picture in one-dimensional systems is not too surprising
owing to the triviality of percolation in $d=1$. This signals that there is
very likely an intermediate dimension below which the universality class that we
have described shifts to some other class. It could even be that FES and PCP
systems behave differently in low space dimensions. As usual with
expansions in the vicinity of the upper critical dimension, few results lend
themselves to direct comparison with the simulations. We shall not try to review
the numerical status of the systems considered throughout this paper, but we
mention that the general belief is that the upper critical dimension, both for
the FES and PCP-like systems is $d_c=4$, and not $d_c=6$ as advocated here. We
refer the interested reader to \cite{lubecketal}.\\

We would like to mention a list of interesting problems. In all three families
of processes that we have considered there is competition between a one-particle
branching process and nonlinear growth limiting processes. It was recently
argued that binary spreading processes (first introduced by~\cite{pcpd1st}), that is reactions in which branching
only occurs by pairs, would lead absorbing state transitions to a universality class
different from that of directed percolation. While there is not even any
consensus on that, we speculate that it is indeed so~\cite{olivieruwefred}, at
least in the case of the Pair Contact Process with Diffusion (PCPD). In
the PCPD there are two absorbing states (degenerate in the large size limit),
but coupling it to an auxiliary field, possibly conserved and/or static, might reveal
a rich variety of behaviors~\cite{chatekockelkoren2}. We leave this and other
issues for future work.

\noindent {\bf Ackowledgment} The research presented here arose from direct collaborations and evergoing, and
sometimes disagreeing, discussions
with H.J. Hilhorst, K. Oerding, A. Vespignani, H. Chat\'e and M.A. Mu\~noz, who
are all gratefully acknowledged. The author thanks the Fundamenteel Onderzoek
der Materie (FOM) and the Lorentz Fonds for financial support.


\end{document}